\newcommand{\eqb}{\begin{equation}}
\newcommand{\eqe}{\end{equation}}
\documentclass[pre,aps,twocolumn,superscriptaddress,amsmath,showpacs,floatfix]{revtex4}
\usepackage{graphicx}

\begin{document}

\title{The long reach of DNA sequence heterogeneity in diffusive processes}
\author{Michael Slutsky}
\email{mich@mit.edu}
\affiliation{Department of Physics, Massachusetts Institute of
Technology, 77 Massachusetts Avenue, Cambridge, MA 02139, USA}

\author{Mehran Kardar}
\affiliation{Department of Physics, Massachusetts Institute of
Technology, 77 Massachusetts Avenue, Cambridge, MA 02139, USA}

\author{Leonid A. Mirny}
\affiliation{Department of Physics, Massachusetts Institute of
Technology, 77 Massachusetts Avenue, Cambridge, MA 02139, USA}
\affiliation{Harvard-MIT Division of Health
Sciences and Technology, Massachusetts Institute of Technology, 
77 Massachusetts Avenue, Cambridge, MA 02139, USA}

\begin{abstract}

Many biological processes involve one dimensional diffusion over 
a correlated inhomogeneous energy landscape  with a correlation length $\xi_c$.
Typical examples are specific protein target location on DNA, 
nucleosome repositioning, or DNA translocation through a nanopore,
in all cases with $\xi_c\approx$ 10~nm.
We investigate such transport processes by the mean first passage time
(MFPT) formalism, and find diffusion times which exhibit strong
sample to sample fluctuations. 
For a a displacement $N$, the average MFPT is diffusive, while its
standard deviation over the ensemble of energy profiles scales as
$N^{3/2}$ with a large prefactor.
Fluctuations are thus dominant for displacements smaller than a
characteristic $N_c \gg \xi_c$: typical values are much less than the
mean, and governed by an anomalous diffusion rule. 
Potential biological consequences of such random walks, composed
of rapid scans in the vicinity of favorable energy valleys and occasional
jumps to further valleys, is discussed.

\end{abstract}

\maketitle
\section{Introduction}

Diffusion appears in most basic processes in the living matter and
therefore has been studied extensively by theoretical and experimental
biophysicists for many decades. At the
macroscopic scale, the phenomena are adequately described by continuum
models that form a well established methodology finding many
applications in science and technology
\cite{murray}. Advanced experimental methods, such as nanoprobing and
single-molecule techniques, provide us with a wealth of data at the
microscopic level. Theoretical description of the observed phenomena
at such scales
is often a considerable challenge, since many irregular features that
average out on the macroscopic scale cannot be ignored
anymore. Sometimes, however, rather simple characteristics
emerge, allowing for exact analytic treatment.

One-dimensional (1D) transport is rarely found on the macroscopic
scale; at the molecular level though, one can find several examples,
e.g. kinesin motion along microtubules \cite{motors, vale, astumian}
or DNA translocation through a nanopore \cite{lub_nels, degennes_dna,
elbaum, meller}.  Usually, in such problems, the underlying potential
profile is considered to be constant or at least regular. However, as
we show in this paper, DNA sequence heterogeneity and the resulting
random energy landscape can have a considerable influence on the
diffusion up to biologically relevant length scales at room
temperatures.

\subsection{Protein-DNA interaction}

\begin{figure}[htb]
\includegraphics[width = 2.7in]{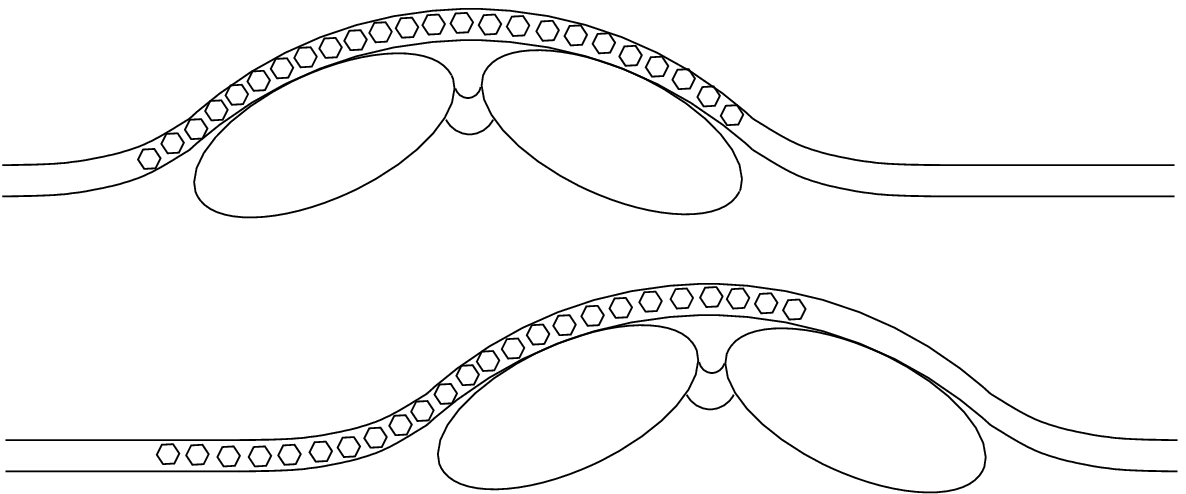} \\
(a)\\
\includegraphics[width = 2.7in]{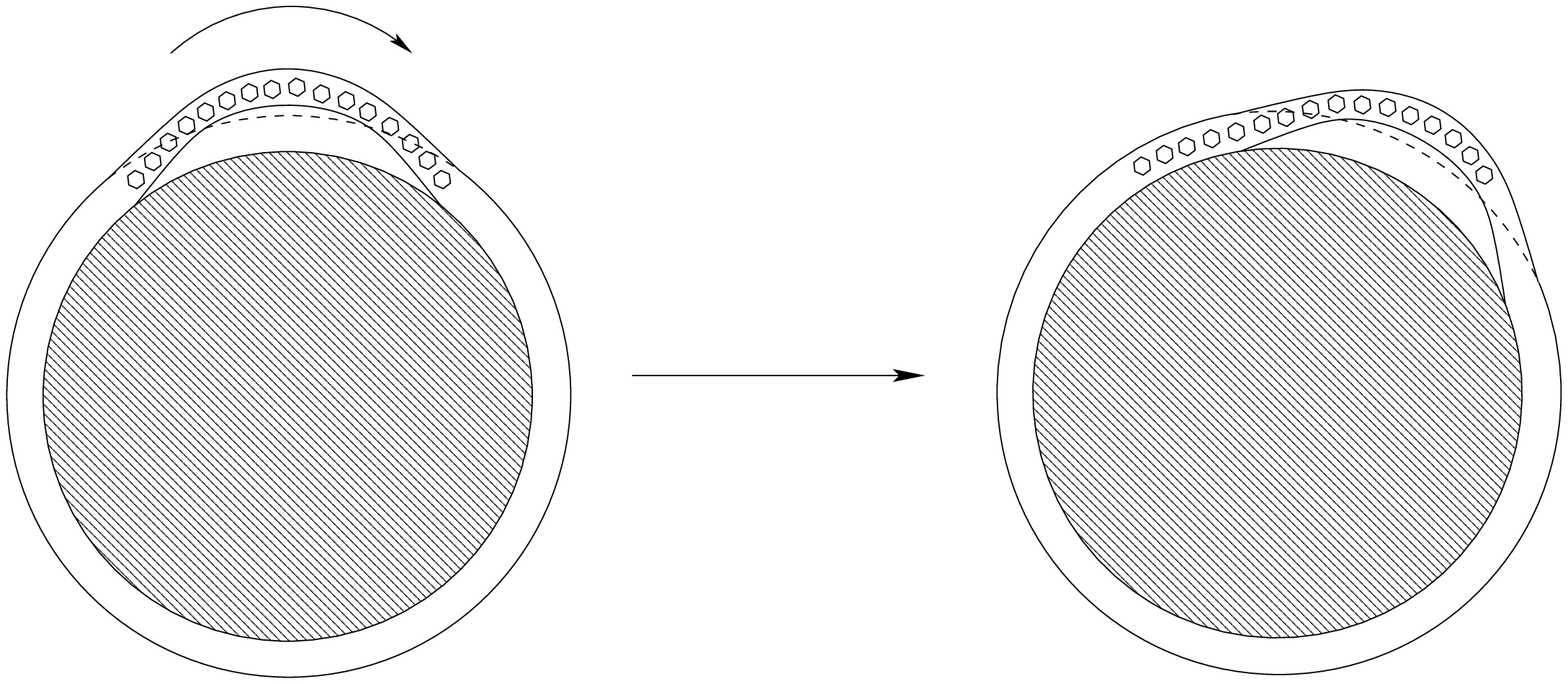}\\
(b)
\caption{\label{fig:bio_ex} (a) Prokaryotic transcription factor sliding;
(b) Nucleosome repositioning.}
\end{figure}

The first example we study here arises in the context of protein-DNA
interaction. As proposed by von Hippel and Berg
\cite{bvh_kin1, bvh_kin2}, and recently observed in many systems
\cite{shimamoto}, 1D ``sliding'' of proteins along the DNA
molecule is an important component of protein specific site location;
at least in prokaryotes.  The ``sliding'' is viewed as an unbiased,
thermally activated process. The actual rules of motion for sliding
depend on the details of interaction between the protein and the
DNA. The general belief is that there are two protein-DNA binding
modes: a strong ``specific'' mode that characterizes binding of
operator sites, and a much weaker ``non-specific'' mode in which
binding of non-cognate DNA occurs \cite{bvh_kin2, bvh1, hwa,
bruinsma}.  In the ``non-specific'' or ``search'' mode, the
interaction energy is usually assumed to be independent of the DNA
sequence that the protein is bound to, though not much experimental
evidence beside relatively fast observed search times favors this
strictly ``equipotential'' picture.  On the other hand, scanning force
microscopy experiments by Erie et. al. \cite{bustamante} clearly
demonstrate DNA bending by {\it Cro} repressor protein, both at
operator and at non-operator sequences
\footnote{DNA bending by transcription factors is a well-known
phenomenon, though practically all the available experimental data
focus on proteins bound to operator sequences.}. Since local DNA
elasticity is known to be highly sequence-dependent \cite{dna_bend},
the energy of protein bound at random locations should have a random
component, correlated at length scales of the order of the protein
binding domain size; see Fig.~\ref{fig:bio_ex}a. This
sequence-dependent interaction energy component appears in addition to
possible local uncorrelated sequence-dependent contributions
from amino acid - base pair contacts.

\begin{figure}[htb]
\begin{tabular}{c c}
\includegraphics[width = 1.6in]{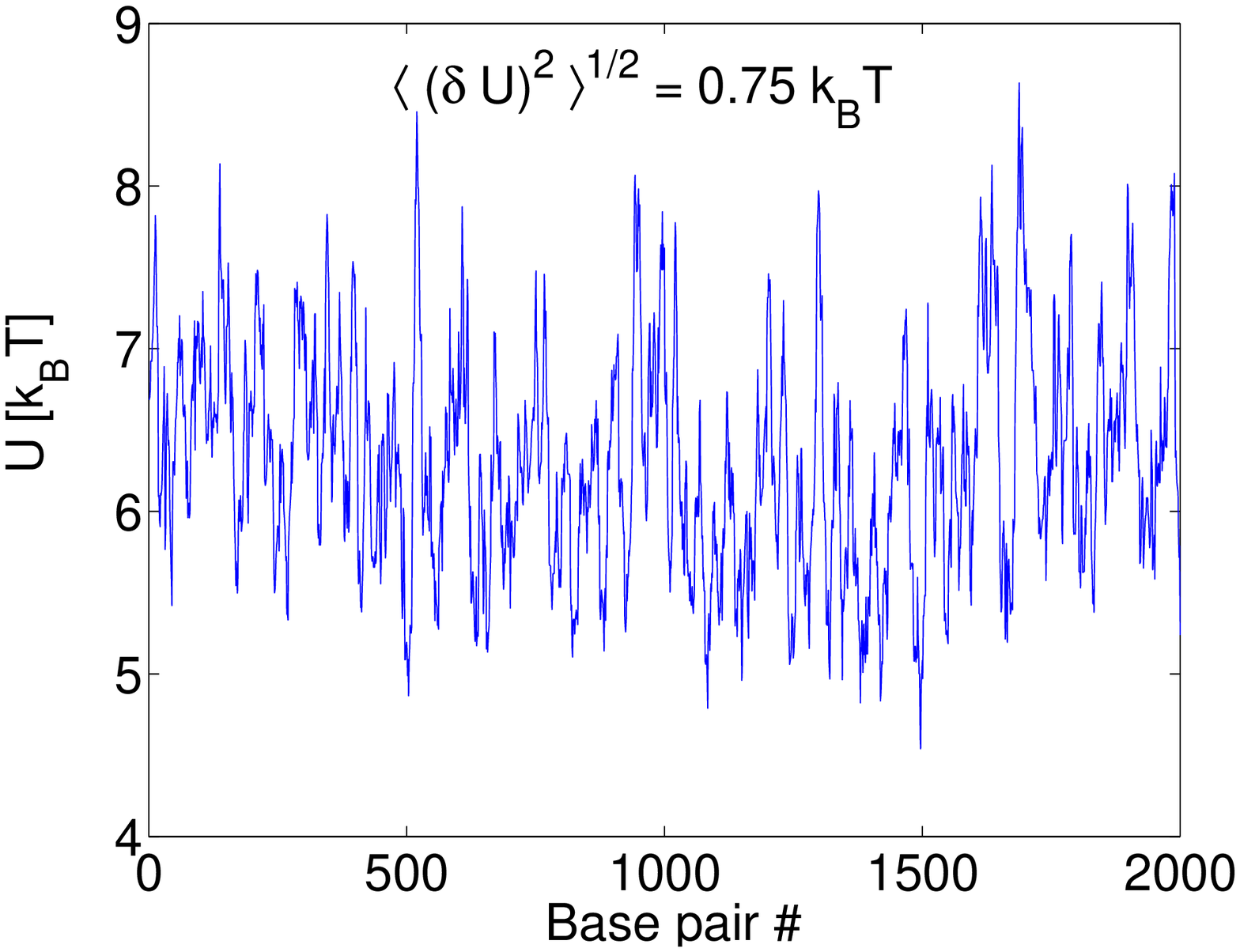} &
\includegraphics[width = 1.6in]{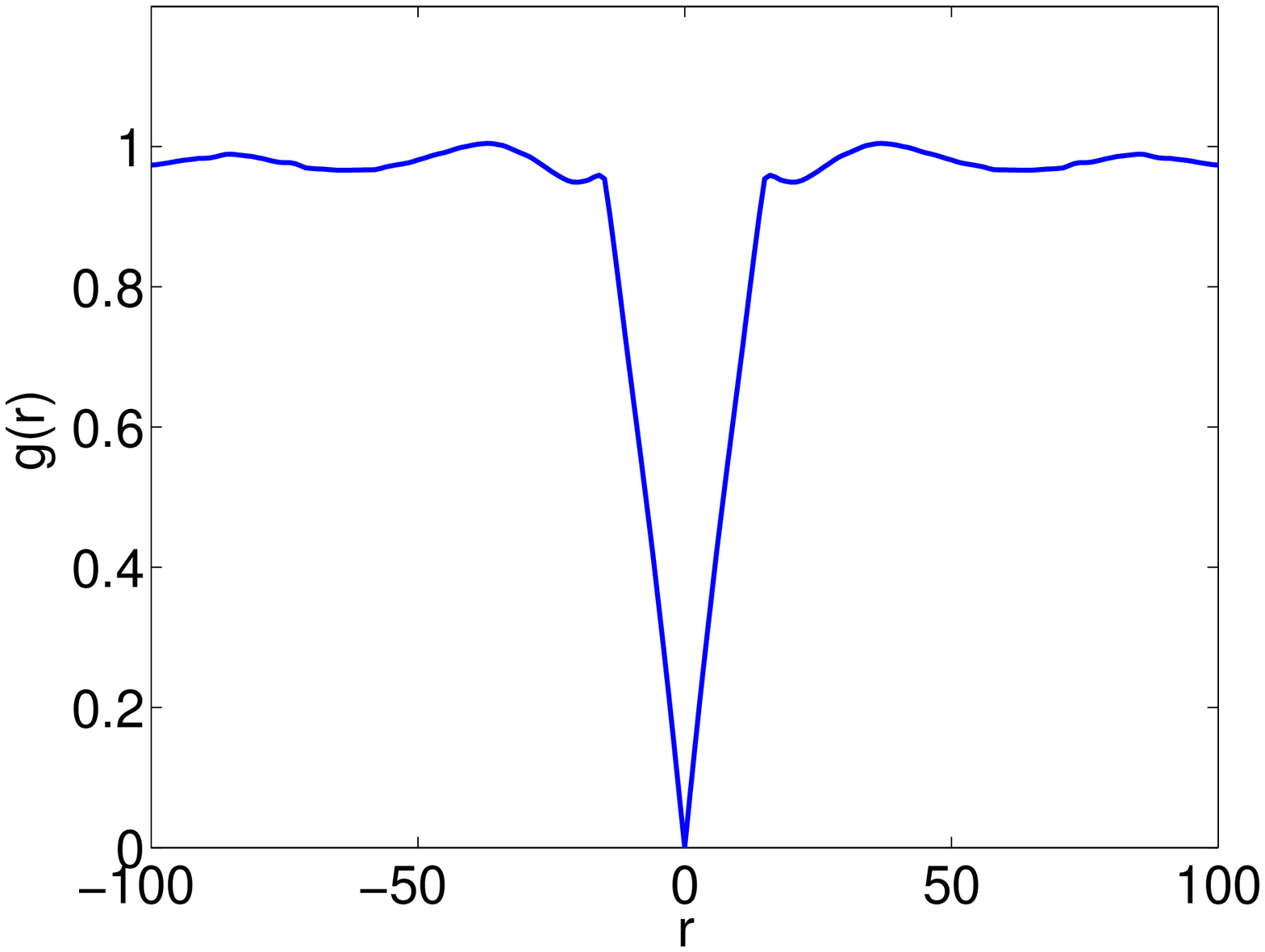}\\
(a)& (b)\\
\end{tabular}
\caption{\label{fig:elast} 
 (a) Energy of local elastic deformation and (b) Potential profile correlator,
as calculated from the data supplied by the server BEND.IT for a
segment of $E. ~coli$ genome. 
The deformed DNA sequence is assumed to be of
length  $L =15~\rm{bp}$.}
\end{figure}

To estimate the significance of the random component of the elastic
energy, we use DNA elasticity data supplied by the BEND.IT server
\cite{vlahovicek}, that incorporates DNase I based bendability
parameters \cite{brukner} and the consensus bendability scale
\cite{gabrielian}. We assume that the protein-DNA complex in
Fig.~\ref{fig:bio_ex}a has a fixed geometry, i.e. the protein is
``hard.'' Then, the random component of the binding energy $\delta U$
is proportional to the random component of the Young's modulus $\delta
E$
\eqb
\delta U = \frac{\delta E}{\bar E} \left(\frac{\ell_p \theta^2}{2L}\right)k_BT,
\eqe
where $\ell_p\simeq 50~\rm{nm}$ is the DNA persistence length,
$\theta\simeq 60^\circ$ is the curvature angle \cite{bustamante}, $L =
10-20~\rm{bp}$ is the bent sequence length and ${\bar E}\simeq
3.4\times 10^{8}~\rm{N/m}$ is the average Young's modulus. The
resulting potential profile is plotted in Fig.~\ref{fig:elast}a.  The
standard deviation of the random component is $\langle(\delta
U)^2\rangle^{1/2} \sim 0.5-1.5 ~k_BT$, so that disorder appears to be
relevant for this problem.

Another interesting example, also from the field of protein-DNA
interaction, was considered recently by Schiessel et. al. \cite{nucl},
and deals with nucleosome repositioning by DNA reptation. It was
argued that chromatin remodeling \cite{widom1, widom2} can be readily
understood in terms of intranucleosomal loop diffusion, the size of
the loop resulting mainly from a compromise between elastic energy
and nucleosome-DNA binding energy. Here again, for a given size of the
loop, the elastic energy is sequence-dependent
\cite{widom2}, and therefore has a random component with finite
correlation length; see Fig.~\ref{fig:bio_ex}b.  For nucleosome
repositioning, this effect may be even more pronounced than for
prokaryotic protein-DNA interaction; the bending angles $\theta$
and the sequence lengths $L$ are 2-3 times larger so that the net
effect may be twice as strong as for the $Cro$ repressor \cite{nucl}.

It is known that DNA can have an $intrinsic$ curvature arising from
the stacking interactions between base pairs. Such sequence-dependent
curvature can play a role similar to sequence-dependent DNA
bendability in providing a correlated landscape. The bending energy of
an intrinsically curved region is easier, requiring a smaller angular
deformation $\theta=\theta_{\rm complex}-\theta_{\rm intrinsic}$ by
the DNA-protein complex. Such sequence-dependent intrinsic curvature
was suggested to be involved in positioning
nucleosomes~\cite{trifonov_2002}.

Aside from DNA bendability and curvature, local correlations in
nucelotide composition, known to be present in eukaryotic genomes,
(AT/GC-rich isochores) can result in a correlated landscape of the
protein-DNA binding energy. This effect becomes especially pronounced
when a DNA-binding protein has a strong preference toward a particular
AT/GC composition of its site. However, in this case, variations take
place over much longer scales, and are not quantitatively relevant in
the specific contexts addressed in this paper.

Both above examples can be viewed as specific cases of DNA reptation by
means of a propagating defect (or ``slack'') of a fixed size. Elastic
energy associated with the slack creation is sequence-dependent and
correlated on the scale of the slack size. The propagating defect is
well localized and samples the energies of well-defined subsequent DNA
segments.  As was pointed out by Cule and Hwa \cite{cule},
short-range correlated randomness of this kind has no effect on the
scaling of the reptation time. However, as we show below, the defect
motion itself is strongly influenced by the disorder and has
nontrivial behavior at different length scales.

\subsection{DNA translocation through a nanopore}\label{sec:translocation}

\begin{figure}[htb]
\includegraphics[width = 3.3in]{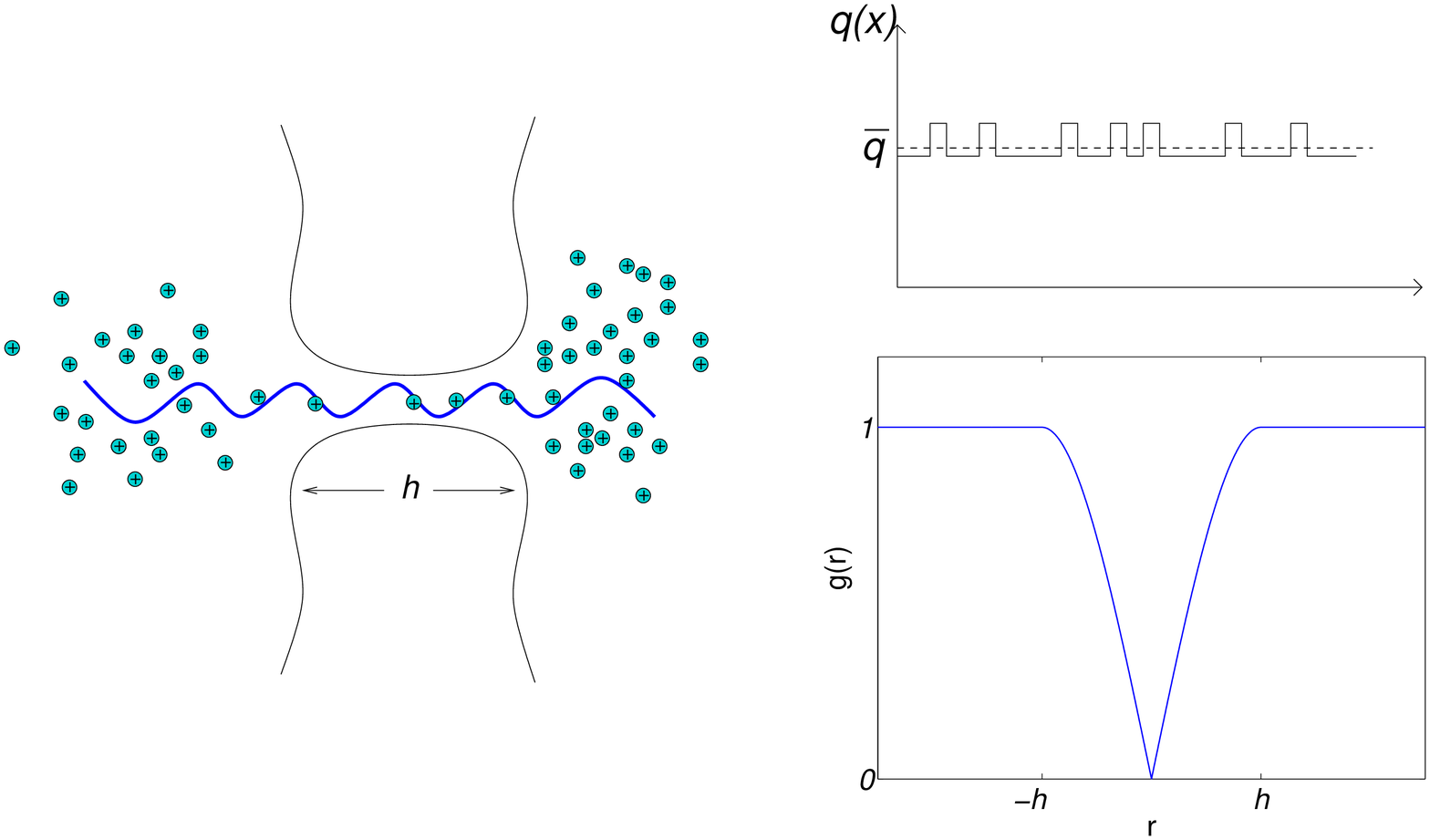}
\caption{\label{fig:pore} ssDNA transport through the nanopore; on the
right: charge density $q(x)$ and correlator  
$g(r) = \langle[\delta U(x) - \delta U(x+r)]^2 \rangle/(2\langle \delta U^2(x)\rangle)$ as a
function of the coordinate $r$.}
\end{figure}

Consider a piece of single-stranded DNA (ssDNA) passing through a large
membrane channel. If the potential difference across the membrane is
zero, the motion of the ssDNA is governed by thermal fluctuations. Since
the channel width differs from the ssDNA external diameter only by few
{\AA}ngstroms \footnote{For $\alpha$-haemolysin, the diameter of the
limiting aperture is about 15 \AA.}, it is reasonable that local
interactions between the nucleotides and the amino acids of the
channel take place. These interactions may have a local base-dependent
component. In addition, longer-range terms are likely to appear in the
presence of a voltage difference. In the cytoplasm, the DNA negative
charge is almost completely screened out at distances of few
nanometers by the counterion cloud. When the DNA molecule enters the
pore, most of the counterions are likely to be ``shaven off,'' though
some of them may remain stuck to the DNA; see Fig.~\ref{fig:pore}.
Thus, the linear charge density inside the pore acquires a random and
basically uncorrelated component:
\eqb
q(x) = \bar{q}(x) + \delta q(x), \qquad 
\langle \delta q(x)~ \delta q(y)\rangle = \rho^2d\delta(x-y).
\eqe
The potential energy of the DNA segment inside the pore in the
presence of a voltage difference of $V_0$ is
\eqb
U(x) = \frac{V_0}{h}\int _{x}^{x+h} x'q(x')dx'.  
\eqe

Since the average charge density $\bar{q}(x)$ is nonzero, DNA
transport is driven by the average force $V_0\bar{q}(x)/h$. The
correlation function of the random component of $U(x)$ is readily
calculated to be
\eqb
\langle \delta U(x)\delta U(x+y) \rangle = 
\frac{V_0^2\rho^2d}{3h^2}(h - |y|)^2\left(h + \frac{|y|}{2}\right)
H(h - |y|),
\eqe
where $H(x)$ is the Heaviside function.  Thus, the potential profile
for DNA motion has a random component with correlation length of
$h$. Taking $V_0 \sim 100~ {\rm mV}$, $\rho \sim e/h$ ($e$ is the
elementary charge), $h \sim 10~ {\rm nm}$, we obtain $\delta U \sim 
k_BT$.

Although this example differs from the above ones in that a nonzero
average driving force is present, large random fluctuations of the
energy landscape may have significant effect on the distribution of
translocation times - a problem that has attracted much interest
lately \cite{meller_review}.

\section{Diffusion in a random potential}

\subsection{The model}

The problems described above map onto a one-dimensional random walk
with position-dependent hopping probabilities $p_i$, $q_i = 1-p_i$ to
the right and to the left, respectively; it is most natural to assume
the regular activated transport form
\eqb\label{eq:prob}
p_i \propto e^{ -\beta\left(U_{i+1} - U_{i}\right)}, \qquad q_i
\propto e^{ -\beta\left(U_{i-1} - U_{i}\right)},
\eqe
where $\beta\equiv \left(k_BT\right)^{-1}$ and $U_i$ is the
sequence-dependent component of the potential energy. The latter is
basically a sum of many random contributions and can therefore be
considered to be normally distributed \cite{hwa}. Thus, in the absence of
correlations, the probability for realization of a certain profile $U(x)$
of length $L$ is (in the continuum limit)
\eqb\label{eq:pdf_uncorr}
P[U(x)]\propto \exp\left[-\alpha\int_0^L dx~U^2(x)\right].
\eqe
This is the well-known Random-Energy Model \cite{rem} that was applied
successfully to various biophysical problems, from protein folding
\cite{wolynes} to protein-DNA interaction \cite{hwa}.  
It assumes no correlations between energies of different
sites. One can think of a more general form of potential profile
\eqb\label{eq:pdf_general}
P[U(x)]\propto \exp\left[-\int_0^L\int_0^L dydx~U(x)G(x-y)U(y)\right].
\eqe
Taking for example, $G(x-y) \propto \partial_{xy}^2\delta(x-y)$, we
obtain the Random-Force Model \cite{bouchaud_ann} that describes an energy
landscape appearing as a random walk with linearly growing
correlations. This model was studied during the last decades in the
context of heteropolymer dynamics \cite{degennes, cule}, glassy
systems \cite{vinokur,fisher} and quite recently - to describe DNA
denaturation dynamics \cite{hwa_bubble}. Characteristic features of
the Random-Force Model are logarithmically slow (``Sinai's'')
diffusion \cite{sinai, goldhirsh1} and aging \cite{fisher,
hwa_bubble}. More generally, $G$ is related to the correllator of $U$
by $\langle U(x)U(y)\rangle = G^{-1}(x-y)$.
 
To include finite-range correlations into Eq.~(\ref{eq:pdf_uncorr}),
we must incorporate a limitation on the acceptable forces. The
ensemble of energy profiles is therefore naturally described by the
following probability density
\begin{subequations}
\eqb\label{eq:pdf}
P[U(x)]\propto e^{-\mathcal{H}[U]},
\eqe
with {\sl pseudoenergy}
\label{eq:pseudo}
\eqb 
\mathcal{H}[U] = \int_0^L dx~\left[\alpha U^2(x) +
\gamma \left(\frac{dU}{dx}\right)^2\right].
\eqe
\end{subequations}
Energy level statistics for this kind of potential profile is also
Gaussian, as can be seen from the average
\eqb\label{eq:char_fun}
\left\langle e^{ikU} \right\rangle = 
\frac{\int\mathcal{D}[U] e^{ikU}
e^{-\mathcal{H}[U]}}
{\int\mathcal{D}[U]e^{-\mathcal{H}[U]}} 
= \exp\left(-{\frac{k^2}{8\sqrt{\alpha\gamma}}}\right),
\eqe
which is the characteristic function for  Gaussian distribution with
zero mean and variance 
\eqb\label{eq:sigma_corr}
\sigma^2 = \frac{1}{4\sqrt{\alpha\gamma}}.
\eqe
The correlator of the potential profile is readily calculated as
\eqb
g(r) \equiv \frac{1}{2}\langle [U(x) - U(x+r)]^2 \rangle
 = \sigma^2 \left(1 - e^{-|r|/\xi_c}\right),
\eqe
where $\xi_c = \sqrt{\gamma/\alpha}$ is the correlation length.

\subsection{Mean First Passage Time}
A convenient formalism for analyzing diffusion in a random
one-dimensional potential profile is that of mean first-passage time
\cite{goldhirsh1, kehr1}.  For a given set of probabilities $\left\{
p_i\right\}$, the mean first-passage time (MFPT) from $i = 0$ to $i =
N$ (in terms of number of steps) is
\eqb\label{eq:mfpt}
\bar{t}_{0,N} = N + \sum _{k = 0}^{N - 1} \omega_k + \sum _{k = 0}^{N - 2} \sum _{i = k + 1}^{N - 1} \left( 1 + \omega_k \right) \prod_{j = k + 1}^{i} \omega_j,
\eqe
where $\omega_i \equiv q_i/p_i$ (see Appendix~\ref{app1} for
derivation). The MFPT given by this expression is for a fixed
realization of probabilities, i.e. for a given potential energy
profile; as such, it is itself a random variable. The
disorder-averaged version of the MFPT is readily obtained after
we note that the sequential products in Eq.~(\ref{eq:mfpt}) reduce to
\eqb\label{eq:prod}
 \prod_{j = k}^{i} \omega_j = \exp\left[\beta(U_{i+1} + U_i - U_k
- U_{k-1})\right].
\eqe
For an uncorrelated potential profile, this exponential factorizes into
independent exponentials; after the ensemble averaging and
the summations are carried out, we obtain for $N \gg 1$
\eqb\label{eq:diff_law_uncorr}
\langle\bar{t}_{0,N}\rangle = N^2e^{2\beta^2\sigma^2},
\eqe
where, for the uncorrelated potential ($\gamma = 0$)
\eqb
\sigma^2 = \frac{1}{2\alpha d},
\eqe
where $d$ is the lattice spacing.
Note that this expression cannot be obtained by simply putting $\gamma
= 0$ in Eq.~(\ref{eq:sigma_corr}). The reason is that when $\gamma$
becomes small, the discrete nature of the underlying lattice starts to
matter. The integration in the momentum space extends only up to
$|q_{\rm max}| = \pi/d$, and thus,
\eqb
\sigma^2|_{\gamma \to 0 } = 
\int^{\pi/d}_{-\pi/d}\frac{dq}{4\pi\alpha} = \frac{1}{2\alpha d}.
\eqe

Returning to the case of a finite correlation length, we note that in
the limit of $\xi_c\gg d$, variations of the potential between neighboring
sites can be neglected compared to variations between sites separated
by distances of order $\xi_c$ or larger. Since the main contribution
to the MFPT comes from the double sum in Eq.~(\ref{eq:mfpt}), we can write
the continuum version as
\eqb\label{eq:mfpt1}
\bar{t}_{0,N} \simeq 2\int_0^N dx\int_x^N dy~e^{2\beta(U(x) - U(y))}.
\eqe
To average over all possible realizations of
$\left\{U(x)\right\}$, we calculate
\begin{align}\label{eq:corr}
\left\langle e^{2\beta(U(x) - U(y))} \right\rangle = 
\frac{\int\mathcal{D}[U] e^{2\beta(U(y) - U(x))}
e^{-\mathcal{H}[U]}}
{\int\mathcal{D}[U]e^{-\mathcal{H}[U]}} \nonumber \\
 =\exp\left[\frac{\beta^2\xi_c}{\gamma}(1 - e^{-|x-y|/\xi_c})\right].
\end{align}
For $|x-y| \ll \xi_c$, Eq.~(\ref{eq:corr}) reduces to
$\exp(\beta^2|x-y|/\gamma)$, so that for $N \ll \xi_c$ we have
\eqb
\langle\bar{t}_{0,N}\rangle \sim N^2 \exp(4\beta^2\sigma^2 N/\xi_c).
\eqe
(Here and in what follows, we measure distances in units of $d$,
unless specified otherwise.) This kind of exponential creep is quite
expected, since for $\alpha \to 0$, $\xi_c \to \infty$ our model
(\ref{eq:pseudo}) reduces to the Random-Force Model.

In the opposite limit $|x-y| \gg \xi_c$, we can neglect the
exponent $e^{-|x-y|/\xi_c}$ , so that Eq.~(\ref{eq:mfpt1}) produces an
ordinary diffusion law, with a disorder-renormalized diffusion
coefficient:
\eqb\label{eq:diff_law_corr}
\langle\bar{t}_{0,N}\rangle = N^2e^{4\beta^2\sigma^2}.
\eqe

Comparing Eqs. (\ref{eq:diff_law_corr}) and
(\ref{eq:diff_law_uncorr}), we see that diffusion in a correlated
potential profile proceeds more slowly than in an uncorrelated
profile. It is straightforward to obtain an expression for the
disorder-averaged MFPT for arbitrary correlation length. If we keep
all four terms in the exponential in Eq.~(\ref{eq:prod}) while going
to the continuum limit, we obtain
\eqb\label{eq:mfpt3}
\bar{t}_{0,N} \simeq 
2\int_0^N dx\int_x^N dy~e^{\beta(U(x+d)+ U(x) - U(y) - U(y-d))}.
\eqe
Averaging this expression over the disorder as in Eq.~(\ref{eq:corr})
yields for $N \gg \xi_c$
\eqb\label{eq:diff_law_general}
\langle\bar{t}_{0,N}\rangle = N^2\exp[2\beta^2\sigma^2(1 + e^{-d/\xi_c})],
\eqe
which has the obvious limits of Eqs.~(\ref{eq:diff_law_uncorr}) and
(\ref{eq:diff_law_corr}) for $\xi_c \to 0$ and $\xi_c \gg d$,
respectively.

\section{Typical vs average}
	
Large deviations from the average are characteristic to many
disordered systems. In this section, we therefore explore the {\sl
typical} properties of random walks as compared to the
disorder-averaged ones.

\subsection{Quantifying fluctuations}

\begin{figure}[hb]
\begin{tabular}{cc}
\includegraphics[width = 1.6in]{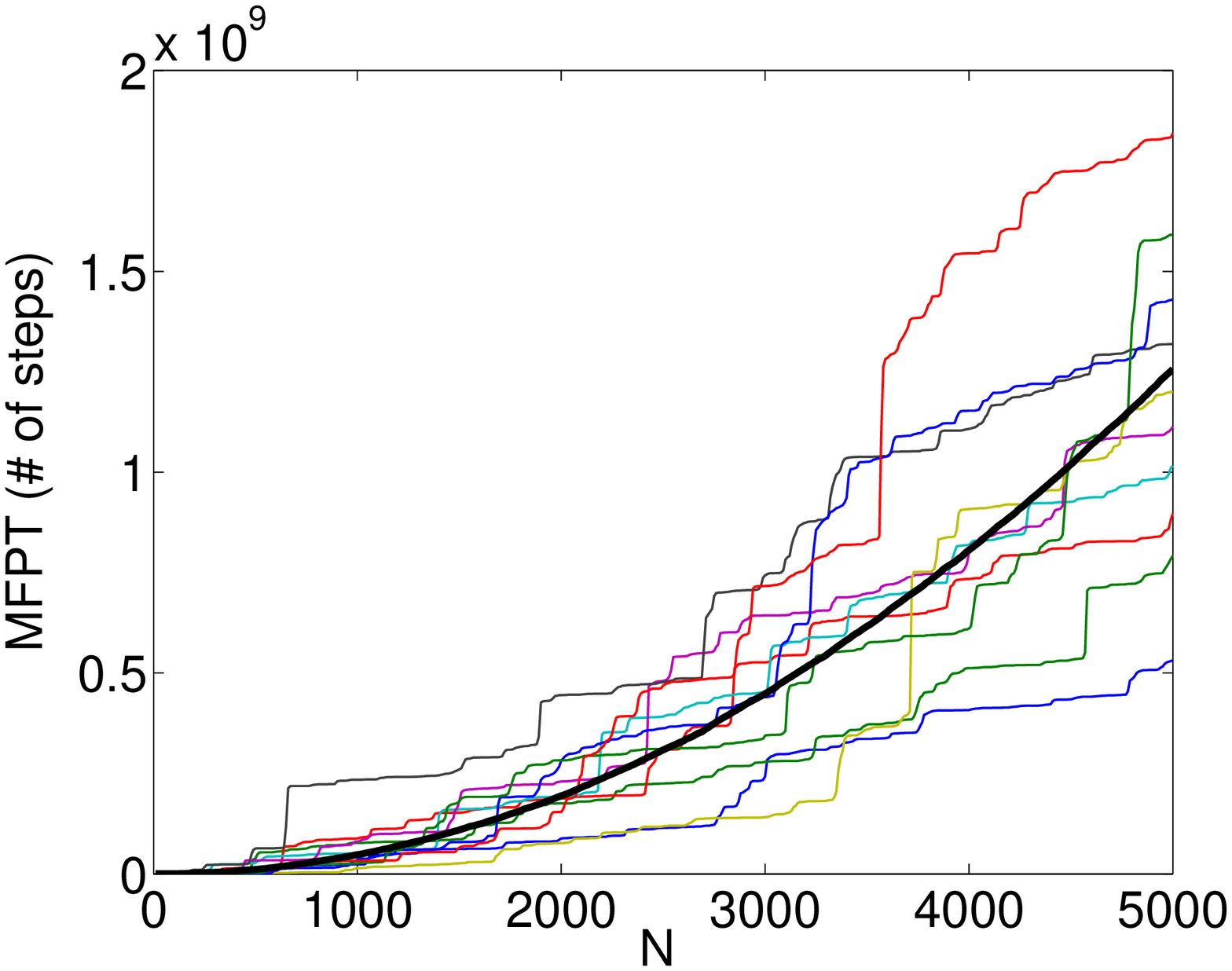}  &
\includegraphics[width = 1.6in]{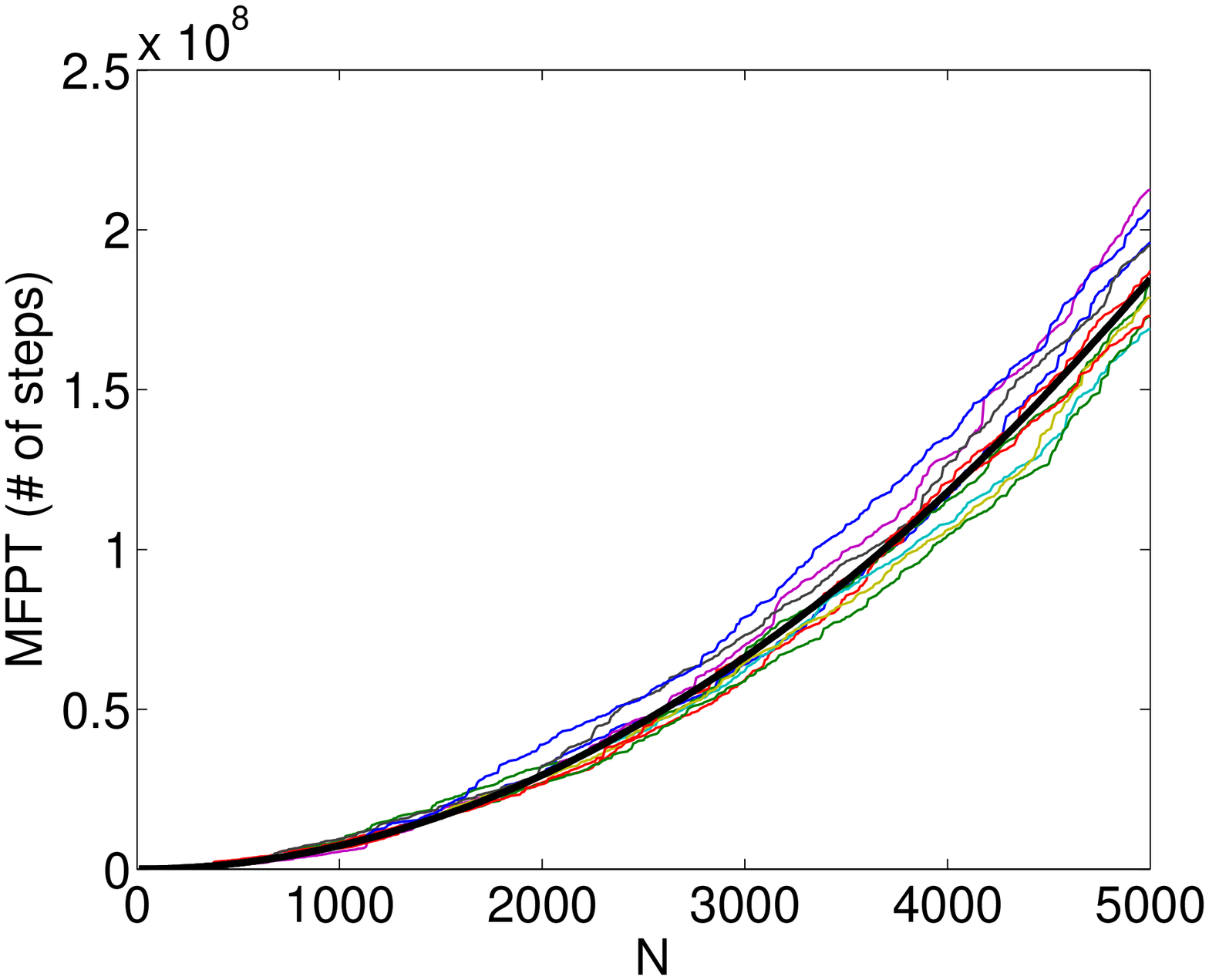} \\
(a)  & (b)
\end{tabular}
\caption{\label{fig:mttf}  Mean First Passage Times: typical versus
average. Thick solid lines are the result of averaging over 1000
realizations of potential profiles ($\beta\sigma = 1.0$): (a)
correlated profile with $\xi_c = 40.0$; (b)
uncorrelated profile.}
\end{figure}

\begin{figure}[htb]
\includegraphics[width = 2.7 in]{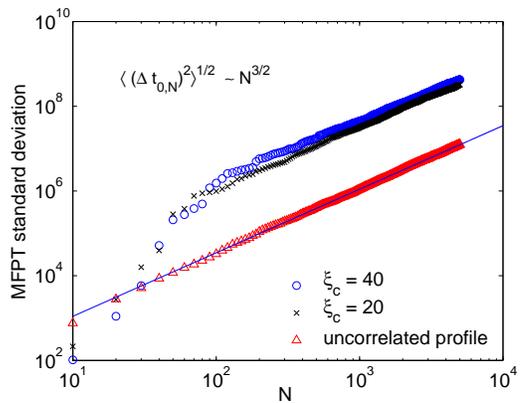}
\caption{\label{fig:std} MFPT standard deviation for $\beta\sigma =
1.0$ for correlated and uncorrelated potential profiles.}
\end{figure}

After the potential profile is generated (see Appendix \ref{app2}), we
calculate the MFPT using Eq.~(\ref{eq:mfpt}). Fig.~\ref{fig:mttf}a
presents the mean first passage times calculated for various 
realizations of $U(x)$ at biologically relevant temperature ($\sigma
\simeq k_BT$). It is clear that although the ensemble-averaged MFPT
does behave as prescribed by Eq.~(\ref{eq:diff_law_general}), typical MFPT
exhibits high variability from one profile to another. The stepwise
shape of typical curves suggests that a random walk in such a profile
consists of regions characterized by subdiffusion (vertical ``steps'')
and superdiffusion (plateaus), appearing intermittently.
Uncorrelated potential profiles, as Fig.~\ref{fig:mttf}b shows, also
lead to a certain disorder-induced variability, though of a
considerably smaller magnitude.  To quantify the sample dependence of
the MFPT, we calculate its variance over the ensemble of potential
profiles. Fig.~\ref{fig:std} presents the standard deviation in
$\bar{t}_{0,N}$ as a function of $N$ for correlated as well as
uncorrelated potential profiles. We observe that the variance scales
as $N^3$ for all profiles. This dependence can be obtained
analytically in a quite straightforward fashion. Consider the average
of the square of MFPT in a potential profile with correlation length
$\xi_c$. The leading term is obviously $N^4\exp(8\beta^2\sigma^2)$ and
it comes from independent $\{i,k,l,m\}$. The next largest contribution
comes from terms with $i = m$ or $k = l$. There are $\sim N^3$ of such
terms, each contributing $\exp(12\beta^2\sigma^2)$. Next, we note that
in order to make a contribution of the same order of magnitude, the
two indexes ($i,m$ or $k,l$) should not necessarily coincide exactly;
it is sufficient that they are less than one correlation length
apart. Hence, after the leading $O(N^4)$ term is cancelled by
$\langle\bar{t}_{0,N}\rangle^2$, the variance is
\eqb
\langle(\Delta\bar{t}_{0,N})^2\rangle \sim
\xi_c N^3 \exp(12\beta^2\sigma^2).
\eqe
Similar reasoning yields for the uncorrelated case
\eqb
\langle(\Delta\bar{t}_{0,N})^2\rangle \sim
{N^3} \exp(6\beta^2\sigma^2).
\eqe
We see that for given $\sigma$ and $\beta$, the correlated energy
landscape produces stronger fluctuations in MFPT than uncorrelated
ones, in agreement with Fig.~\ref{fig:mttf}.

\begin{figure}[htb]
\begin{tabular}{cc}
\includegraphics[width = 1.6in]{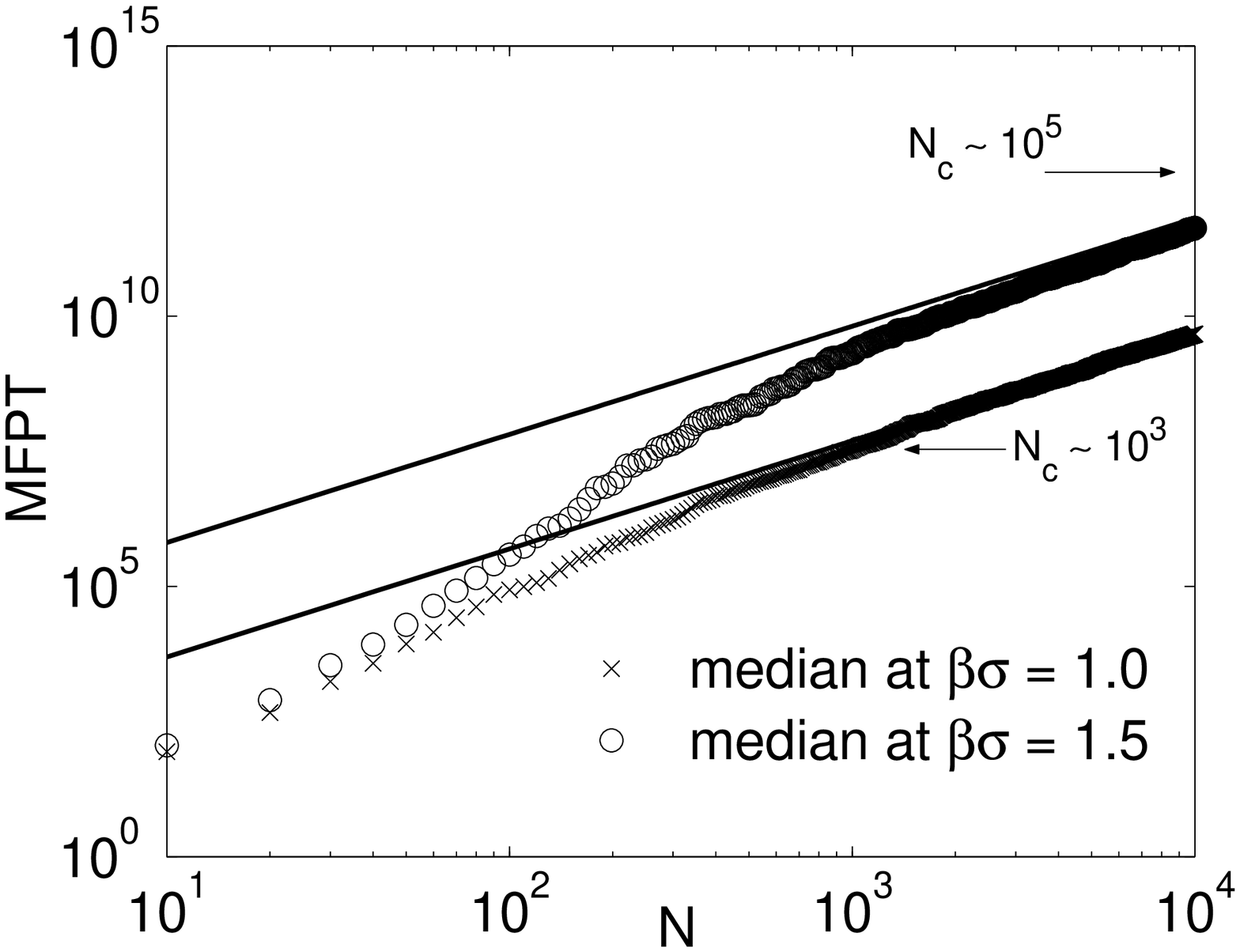}  &
\includegraphics[width = 1.6in]{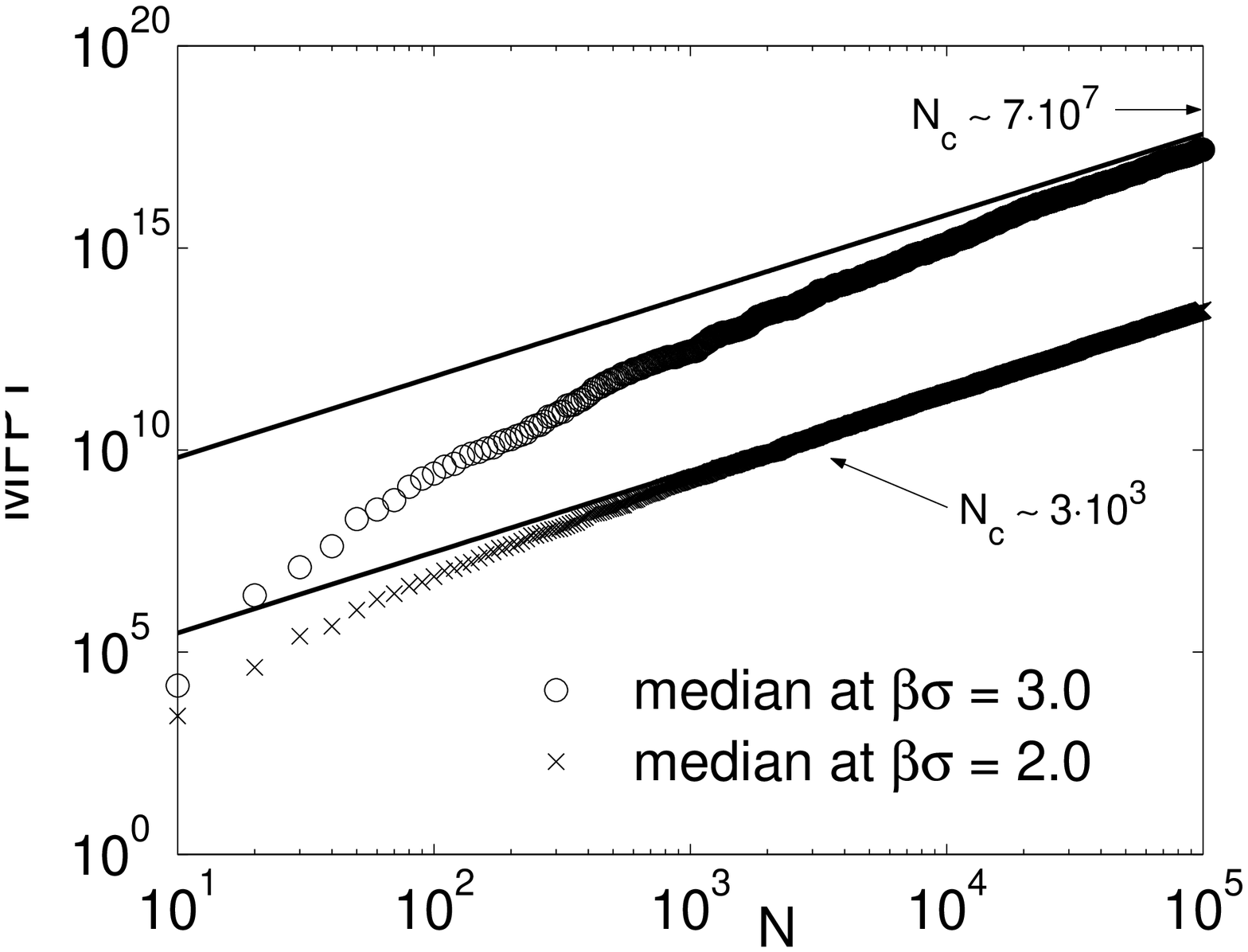} \\
(a)  & (b)
\end{tabular}
\caption{\label{fig:freez} Median versus disorder-averaged (solid lines
calculated from Eq.~(\ref{eq:diff_law_general})) MFPT. Median values
were calculated for 1000 realizations of potential profiles: (a) Correlated potential profile
with $\xi_c = 20.0$; (b) Uncorrelated potential profile.}
\end{figure}

Comparing the expressions for the variance with the corresponding
expressions for disorder-averaged MFPT, we see that for any
temperature, there is a characteristic distance $N_c$, below which
there is no self-averaging and the typical MFPT is determined by
fluctuations. This length is
\eqb
N_c \sim \xi_c e ^{4\beta^2\sigma^2}
\eqe
for correlated profiles, and 
\eqb
N_c \sim e ^{2\beta^2\sigma^2}
\eqe
for uncorrelated ones. This effect is akin to ``freezing'' in the
Random-Energy Model \cite{rem}: for low enough temperatures, typical
passage times for distances below $N_c$ are dominated by high
barriers. This is more pronounced for correlated profiles since in
addition to stronger temperature dependence, there is amplification by
a factor of $\sim\xi_c$, as sites within a correlation length give
similar contributions. Figure~\ref{fig:freez} demonstrates the lack of
self-averaging for uncorrelated potential profiles at short distances
and low temperatures: the {\sl median} MFPT (defined as the 50th
percentile of a sample) shows large deviations from the average at
distances shorter than $N_c$ and coincides with it at distances larger
than $N_c$.

\subsection{Anomalous diffusion}
The lack of self-averaging in the region $\xi_c \ll N \ll N_c$ can be
quantified by estimating the typical MFPT. Consider
Eq.~(\ref{eq:prod}) for an uncorrelated potential and define the
following coarsening procedure: 
$\widetilde{U}_i = U_{2i} + U_{2i+1}$. Then, in the ``freezing regime,'' the double sum
\eqb
\sum_k\sum_i \exp[\beta(\widetilde{U}_i - \widetilde{U}_k)],
\eqe
is dominated by $(i,k)$ producing the largest exponent. For a finite
sample $\{U_i\}$ of size $N$ and variance $\sigma^2$, the
corresponding sample $\{\widetilde{U}_i\}$ contains $N/2$ values
distributed with a variance $2\sigma^2$. The minimum and the maximum
of  $\{\widetilde{U}_i\}$ have therefore characteristic values of $\pm
2\sigma\sqrt{\ln[{N}/({2\sqrt{2\pi}})]}$, respectively. Thus, a typical
MFPT for an uncorrelated potential reads
\eqb\label{eq:typ_uncorr}
\bar{t}_{0,N}\sim \exp\left[4\beta\sigma \sqrt{\ln \frac {N}{2\sqrt{2\pi}}} \right].
\eqe
For the purposes of estimating the extreme values of a correlated energy
landscape, the sample size is effectively reduced by a factor of $\sim
\xi_c$, therefore, the extrema of $\{U_i\}$ are approximately
$\pm \sigma\sqrt{2\ln[{N}/({\xi_c\sqrt{2\pi}})]}$. Noting that sites within
a correlation length around the extrema contribute similarly to the
MFPT, for a correlated potential we write 
\eqb\label{eq:typ_corr}
\bar{t}_{0,N}\sim 
\xi_c^2\exp\left[4\beta\sigma \sqrt{2\ln \frac {N}{\xi_c\sqrt{2\pi}}} \right].
\eqe
\begin{figure}[htb]
\begin{tabular}{cc}
\includegraphics[width = 1.6in]{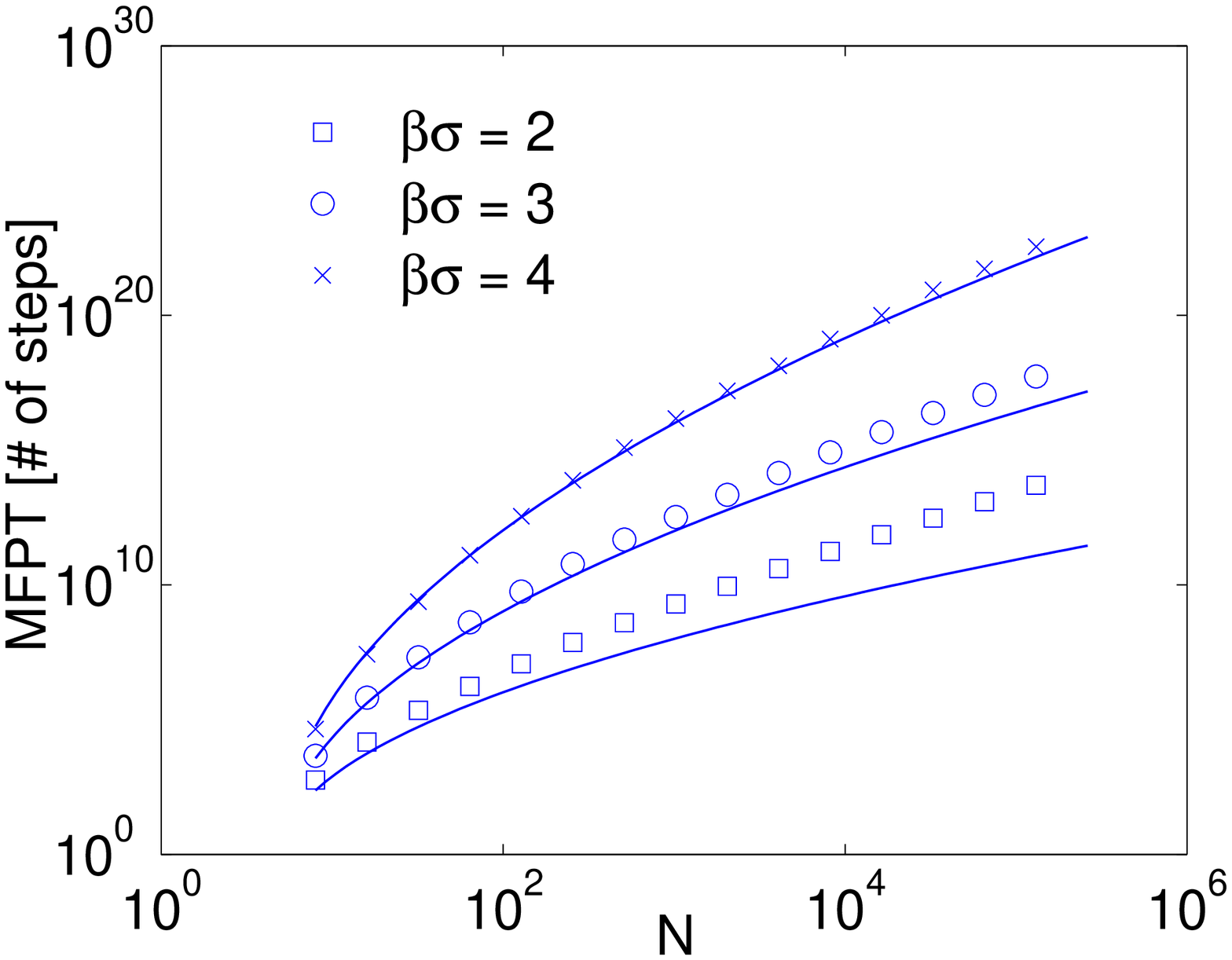}  &
\includegraphics[width = 1.6in]{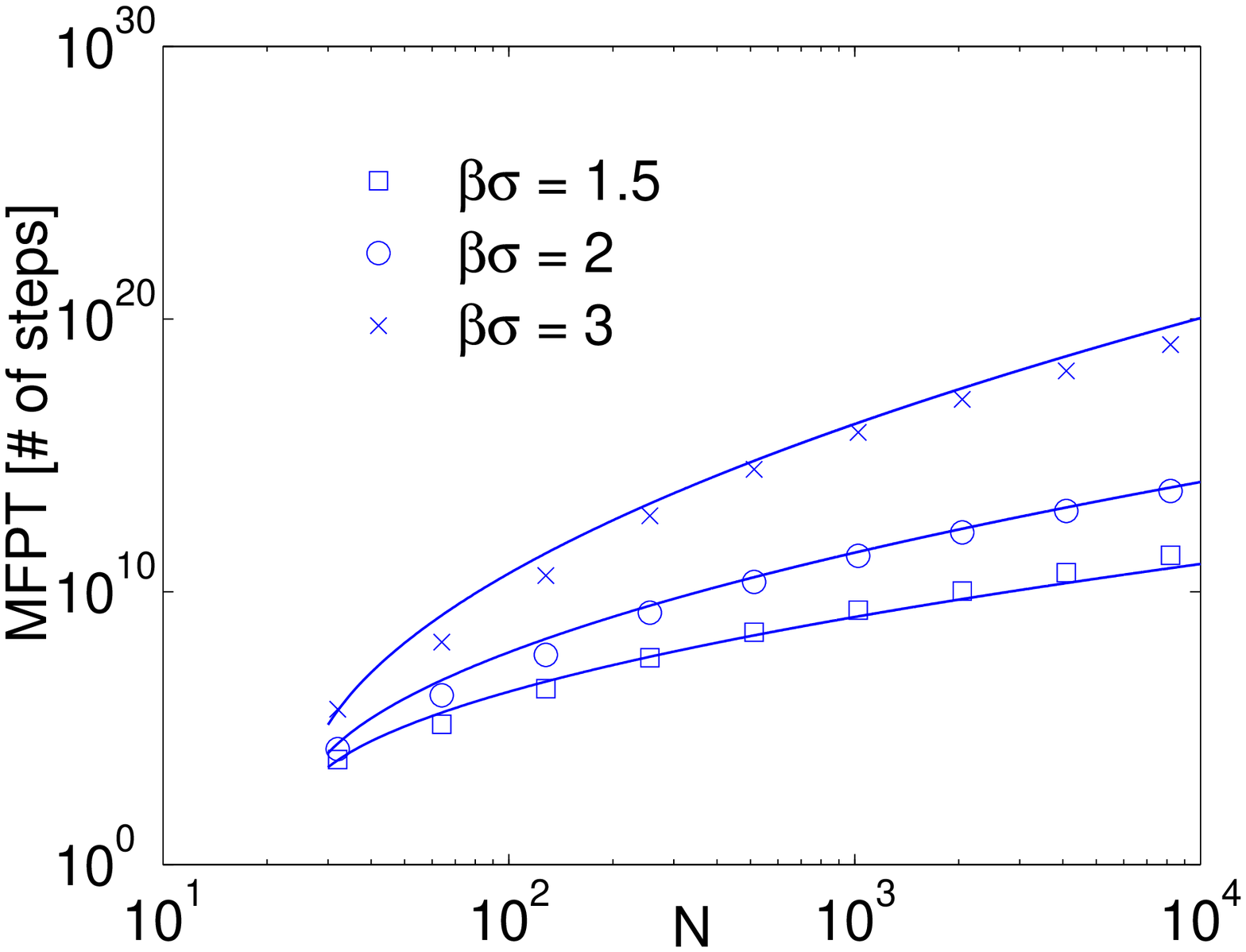} \\
(a)  & (b)
\end{tabular} 
\caption{\label{fig:freez_corr} Typical MFPT for $N \ll N_c$ at various values of $\beta\sigma$: (a) Uncorrelated potential profile; (b) Correlated potential profile with $\xi_c = 10$. Solid lines are the analytical estimates from Eqs.~(\ref{eq:typ_uncorr}) and (\ref{eq:typ_corr}).}
\end{figure}
Figure~\ref{fig:freez_corr} compares typical values of $\bar{t}_{0,N}$
calculated from Eqs.~(\ref{eq:typ_uncorr}) and (\ref{eq:typ_corr})
with numerically calculated median values of MFPT. We see that our
analytical estimates produce a correct order of magnitude for
$\bar{t}_{0,N}$. As expected, for uncorrelated profiles, the agreement
is better at lower temperatures; for higher temperatures,
Eq.~(\ref{eq:typ_uncorr}) is an underestimation since we do not
include contributions from second-lowest, second-highest, etc., energy
levels. Eq.~(\ref{eq:typ_corr}), on the other hand, turns out to be a
slight overestimation, since we have replaced the average of $\sim \xi_c^2$
terms by their maximum value.

\begin{figure}[htb]
\includegraphics[width = 3.0in]{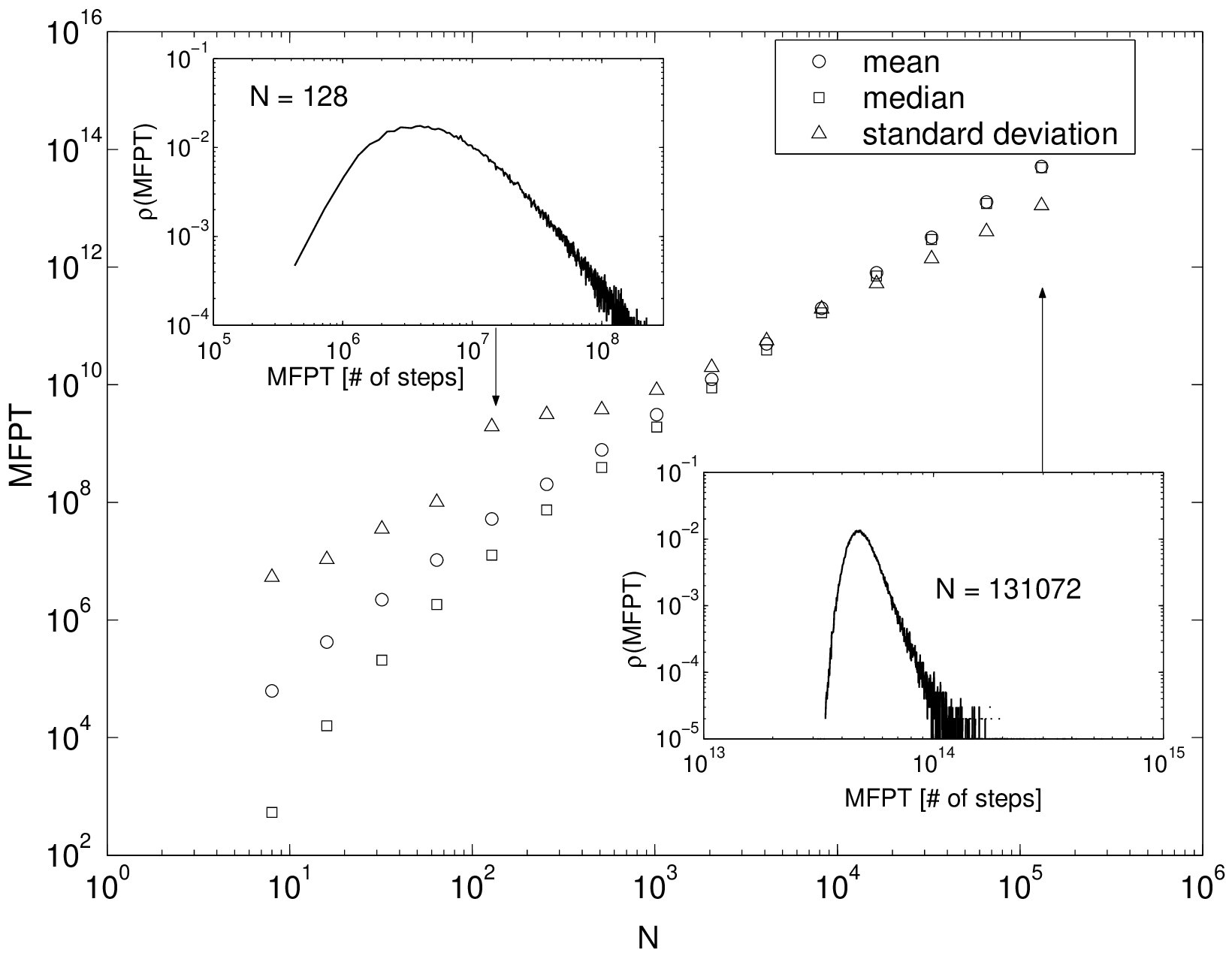}
\caption{\label{fig:distr} Probability density functions for MFPT
calculated for 100,000 uncorrelated profile realizations at
$\beta\sigma = 2$. }
\end{figure}

Large difference between the median and the average values is a
signature of a broad probability distribution. The insets of Fig.~\ref{fig:distr}
present two probability density functions for MFPT, at $N \ll
N_c$ and $N \gg N_c$. For the short distance, the distribution is very
broad and spans several orders of magnitude. For $N \gg N_c$, the
system is self-averaging, in the sense that the MFPT distribution is
much narrower with almost coinciding median and average values.

\subsection{Characteristics of random walk}

\begin{figure}[htb]
\includegraphics[width = 2.7in]{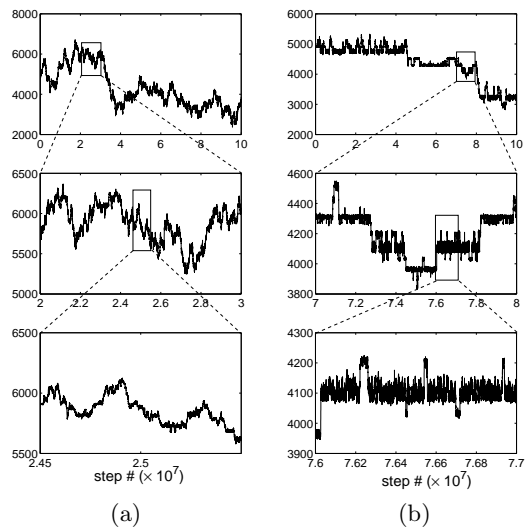}\\
(a) \hskip 1.3in  (b) 
\caption{\label{fig:2rw} Random walk in (a) uncorrelated,  and (b)
correlated with $\xi_c = 20.0$,  potential energy profiles.}
\end{figure}

To complete the picture, we perform direct simulations of random walks
in correlated and uncorrelated potential profiles; typical results are
depicted in Fig.~\ref{fig:2rw}. One can see a clear qualitative
difference between the two cases: random walks in the uncorrelated profile
look very much like  standard walks with $p_i = q_i = 1/2$,
whereas motion of a particle in a correlated profile has a somewhat
different nature.

As above, we see that macroscopic motion of a particle in a correlated
potential consists of subdiffusive as well as superdiffusive
segments. It also appears that the particle tends to be localized near
the bottom of ``valleys'' of few $\xi_c$ in extent,
whereas in an uncorrelated profile, there are no preferable sites for
localization. Obviously, when the time is measured in real-time units,
rather than in number of steps, the particle is more likely to be
found at the minima of the energy landscape in both cases. In terms of the
number of steps though, all sites of the uncorrelated landscape are
revisited more or less uniformly.

\section{Biological implications}

\subsection{Transcription Factors}

Consider a DNA-binding protein searching for its target
site on the genome. As explained in the introduction, a
correlated random energy landscape can arise from the interplay of 
sequence-dependent flexibility, and the bending contribution
to the total DNA-binding energy. 
Diffusion on such a landscape may then lead to localization in
the energy `valleys,' i.e. the protein will reside preferentially
in specific (favorable) areas of the genome.
Such nonuniform sampling has important implications 
for biological strategies of transcription factor bindings:
First, if a valley contains several binding sites, the rapid (superdiffusive)
scanning of the valley leads to quick equilibration between these sites
(while equilibration for similarly spaced sites outside a valley will 
take much longer).
This is important when the protein binds nearby sites with distinct binding 
energies, and the strongest one has to be occupied first to provide correct 
regulation (as in the case of the cro repressor). 
Second, several proteins bind their specific sites
only when activated by ligands (e.g. $PurR$, $GalS$ etc),
spending the rest of the time in an inactive form ``waiting'' for the ligand. 
These proteins can benefit from staying close the site in the waiting mode,
since they can then quickly find their target upon activation. 

One of the results of this study was that inhomogeneities significantly reduce
the overall diffusion rate, as in Eq.~(\ref{eq:diff_law_general}).
While this may be beneficial in confining a protein to favorable regions,
it severely restricts the ability to search large portions of the genome by
one dimensional diffusion.
Since we argue that a portion of the inhomogeneity originates from
variations in the bending energy of the DNA, a potential strategy is
for the binding protein to switch between two states which bend the
DNA weakly or strongly.
The weak bending state is subject to reduced variations in the energy
landscape and can diffusive more freely (search mode), 
compared to the strongly bending state which is more likely to be 
confined in the vicinity of favorable energy valleys (waiting mode).
One potential candidate for exploiting this strategy is the tertarmeric 
$LacI$ protein that consists of two DNA-binding dimeric subunits. 
Each subunit binds DNA and bends it slightly; when both subunits are bound,
DNA is deformed into an extended loop. 
Several experimental results suggest that $LacI$ binds DNA with only 
one subunit while searching for its target site (``holding DNA with one arm''). 
Only when both both subunits find their site, the DNA is bent into a loop. 
Very few structural data are available for proteins bind to DNA non-specifically 
(search mode). The above strategy suggests that DNA is less deformed 
in such complexes.

Another potential source for a correlated inhomogeneous energy landscape is 
an extended protein--DNA interface with net interactions that are the
sum of several local contributions.
(The addition of such correlated contributions leads to a much larger
variance of energy than if they were uncorrelated.)
This can be a significant effect for large multi-protein complexes (such as
polymerases, TFIID, TFIIB complexes in yeast, etc.). 
To avoid slow-down by such inhomogeneities, protein complexes can avoid
scanning DNA in the fully assembled state when the protein--DNA
interface is extensive. Individual components of the complex can
search for their sites independently, assembling the whole complex
only on the right site. In fact, most of large protein--DNA
complexes follow this strategy of assembly on the site, while many
dimers and tetramers are assembled in the solution.

\subsection{Nucleosomes}

Other implications concern nucleosome positioning and dynamics.
Wrapping of the DNA around these large multi-protein complexes 
is essential for packing DNA in the small volume of the cell nucleus. 
Nucleosomes, however, prevent transcription factors and other proteins
from accessing DNA. To allow a transcription factor to access its target, 
nucleosomes close to that site have to be removed from the DNA 
or re-positioned. While removal of nucleosomes is made by specific 
enzymes that chemically modify them (e.g. by histon methilation), 
re-positioning relies in part on nucleosome mobility. 
In general, nucleosomes have to be (i)
positioned at specified locations, and (ii) be able to move along
the DNA in the vicinity to the initial placement site allowing
access to this region of the DNA.

Nucleosome positioning is determined by specific sequences on the DNA. 
Such sequences are also known to provide DNA flexibility
and/or internal curvature \cite{widom_jmb99,trifonov_2002}. 
As discussed above, local DNA flexibility and curvature create a
correlated energy landscape for binding. We suggest that 
inhomogeneous diffusion on such landscapes is an important
element that provides both (i) preferential positioning of the 
nucleosomes due to DNA flexibility and curvature, and (ii) 
relatively rapid diffusion within the confines of the energy valley. 
Conversely, uncorrelated landscapes cannot achieve both objectives,
since strong nucleosome binding sites prevent local diffusion 
along the DNA, while weak sites are not able to localize these proteins, 
leading to their random placement. 
In fact, experiments \cite{widom_jmb99} have shown that
nucleosome positioning sites are extended and are fairly weak.
Such structure of positioning sites creates an extended valley on
the correlated binding landscape, supporting our hypothesis.

This mechanism can also explain how certain proteins (such as
HMGB) can reposition nucleosomes by binding to the DNA in their proximity.
It has been suggested that such proteins alter the local mechanical
properties of the DNA (such as its flexibility, curvature, or super-coiling) 
leading to repositioning of the nucleosome \cite{travers03}. 
If the nucleosome is indeed preferentially localized by being trapped in 
a valley of the binding landscape, HMGB proteins may well alter the shape 
of the valley (e.g. by shrinking it on one side). 
Mobile nucleosomes, rapidly diffusing within the boundaries
of the valley, will then reposition themselves in the new landscape.

\subsection{Translocation}

In Sec.~\ref{sec:translocation} we described how slow (activated)
passage of ssDNA through a nanopore can be modeled by 
diffusion over a correlated landscape. In particular, we demonstrated
that if there are inhomogeneities in the charge of the DNA {\em inside
the channel}, there will be variations in the potential energy
landscape that are proportional to the applied voltage difference $V$.
There is in fact scant structural information about the reconfigurations
of charges (both free and bound) as DNA passes through a channel.
Examining the variations in the MFPT of DNA as a function of the
applied voltage~\cite{meller}, may provide an indirect probe of any
inhomogeneities in the charge passing through a channel.

\section{Conclusions}

We studied one-dimensional diffusion in a random energy landscape with
short-range correlations. We found that disorder with short
correlation length $\xi_c$ leads to a strong sample dependence of
diffusion characteristics. The diffusive transport is influenced up to
length scales exceeding $\xi_c$ by orders of magnitude. Three
diffusion regimes can be identified:
\begin{enumerate}

\item For distances smaller than the correlation length ($N \ll
\xi_c$), the disorder-averaged Mean First Passage Time (MFPT) is 
\eqb
\langle
\bar{t}_{0,N}\rangle \sim N^2 \exp(4\sigma^2\beta^2N/\xi_c).\nonumber
\eqe
At biologically relevant temperatures, the $N^2$ factor
prevails; however, at low temperatures ($k_BT \lesssim
2\sigma/\sqrt{\xi_c}$), we obtain exponential creep (Sinai's
diffusion).

\item For distances $N$ much larger than the characteristic value
$N_c$, MFPT exhibits some variability from sample to sample. However,
the typical value of the MFPT is given by the disorder-averaged MFPT
\eqb
\langle\bar{t}_{0,N}\rangle = N^2\exp[2\beta^2\sigma^2(1 + e^{-d/\xi_c})].\nonumber
\eqe
The variance of MFPT over the ensemble of potential profile
realizations scales as $N^3$ with distance above $N_c$. The
characteristic distance $N_c$ equals $\xi_c e^{4\beta^2\sigma^2}$ for
correlated profiles and $e^{2\beta^2\sigma^2}$ for uncorrelated ones.

\item In the intermediate case $\xi_c \ll N \ll N_c$, the
disorder-averaged MFPT behaves as described by
Eq.~(\ref{eq:diff_law_general}). However, the MFPT distribution over
the ensemble of profile realizations is much broader below $N_c$ than
above it, as Fig.~\ref{fig:distr} demonstrates. As a result, a typical
sample yields diffusion times orders of magnitude shorter than the
average. This effect can be qualitatively understood in terms of the
Random Energy Model. Below $N_c$, diffusion times are mostly
influenced by high barriers and deep valleys that are at the extrema of
energy landscape histogram. The typical diffusion times are given by
\eqb
\bar{t}_{0,N}\sim \exp\left[4\beta\sigma \sqrt{\ln \frac {N}{2\sqrt{2\pi}}} \right] \nonumber
\eqe
for an uncorrelated profile, and
\eqb
\bar{t}_{0,N}\sim 
\xi_c^2\exp\left[4\beta\sigma \sqrt{2\ln \frac {N}{\xi_c\sqrt{2\pi}}} \right], \nonumber
\eqe
for a correlated one. Above $N_c$, most obstacles to the
particle motion lie in the central region, so that
Eq.~(\ref{eq:diff_law_general}) produces a valid estimation for a
typical diffusion time: the system becomes self-averaging.

\end{enumerate} 

These regimes appear to be relevant for biological systems and provide
qualitative insight into the kinetics of protein-DNA interaction.

\begin{acknowledgments}
This work was supported by the National Science Foundation through
grant No. DMR-01-18213 (M.K.). L.M. is an Alfred P. Sloan Research
Fellow and is also supported by John F. and Virginia B. Taplin
Award. M.S. is supported by NEC research fund.
\end{acknowledgments}

\appendix

\section{Mean First-Passage Time Derivation.} \label{app1}

The mean first passage time (MFPT) from site $\#0$ to site $\#N$ is
defined as the mean number of steps the particle has to make in order
to reach site $\#N$ {\sl for the first time}. The derivation here
follows the one in Ref.~\cite{kehr1}.

Let $P_{i,j}\left(n\right)$ denote the   probability to start at  site
$\#i$  and to  reach  the site $\#j$ in   exactly $n$ steps. Then, for
example,
\eqb\label{eq:rec_1}
P_{i,i+1}\left(n\right) = p_i T_i\left(n - 1\right),
\eqe
where  $T_i\left(n\right)$ is defined as  the probability of returning
to the $i$-th site after $n$ steps {\sl without} stepping to the right
of it. Now,  all the  paths contributing  to  $T_i\left(n -  1\right)$
should start with the  step to the left  and then reach the site $\#i$
in $n-2$ steps,  not  necessarily  for   the first time.   Thus,   the
probability $T_i\left(n - 1\right)$ can be written as
\eqb\label{eq:rec_2}
T_i\left(n       -      1\right)     =       q_i       \sum     _{m,l}
P_{i-1,i}\left(m\right)T_i\left(l\right)\delta_{m+l, n-2}.
\eqe
We now introduce generating functions
\eqb
\tilde P_{i, j}\left(z\right) = \sum_{n=0}^{\infty} z^n~P_{i,j}\left(n\right), \qquad \tilde T_{i}\left(z\right) = \sum_{n=0}^{\infty} z^n~T_{i}\left(n\right). 
\eqe
One can easily show (see e.g. Ref.~\cite{goldhirsh2}) that
\eqb
\tilde P_{0, N}\left(z\right) = \prod_{i = 0}^{N-1} \tilde P_{i, i+1}\left(z\right). 
\eqe
Knowing $\tilde  P_{i,  i+1}\left(z\right)$, one calculates   the MFPT
straightforwardly as
\begin{align}
\bar{t}_{0,N} = \frac{\sum_n n P_{0, N}\left(n\right)}{\sum_n P_{0, N}\left(n\right)} = \left[\frac{d}{dz}\ln \tilde P_{0, N}\left(z\right) \right]_{z = 1} 
\nonumber \\
= \sum_{i = 0}^{N-1}\left[\frac{d}{dz}\ln \tilde P_{i, i+1}\left(z\right) \right]_{z = 1}. 
\end{align}
Using Eqs.~(\ref{eq:rec_1}) and  (\ref{eq:rec_2}), we obtain  the following
recursion relation for $\tilde P_{i, i+1}\left(z\right)$:
\eqb\label{eq:P_i}
\tilde P_{i, i+1}\left(z\right) = \frac{zp_i}{1 - zq_i \tilde P_{i-1, i}\left(z\right) }. 
\eqe
To solve for $\bar{t}_{0,N}$, we must introduce boundary
conditions. Let $p_0 = 1,~q_0 = 0$, which is equivalent to introducing
a reflecting wall at $i = 0$.  This boundary condition clearly
influences the solution for short times and distances.  However, as
numerical simulations suggest, its influence relaxes quite fast, so
that for longer times, the result is clearly independent of the
boundary. The benefit of setting $p_0 = 1$ becomes clear when we
observe that
\eqb
\tilde P_{0, 1}\left(1\right) = 1, \qquad \Rightarrow \qquad \forall~i\qquad \tilde P_{i, i+1}\left(1\right) = 1. 
\eqe
Hence,
\eqb
\bar{t}_{0,N} = \sum_{i = 0}^{N-1} \tilde P_{i, i+1}'\left(1\right). 
\eqe
The recursion relation   for $P_{i,  i+1}'\left(1\right)$  is  readily
obtained from Eq.~(\ref{eq:P_i}):
\eqb
\tilde P_{i, i+1}'\left(1\right) = \frac{1}{p_i} + \frac{q_i}{p_i} \tilde P_{i-1, i}'\left(1\right) = 1 + \omega_i \left[1 + \tilde P_{i-1, i}'\left(1\right)\right],
\eqe
with  $\omega_i     \equiv   q_i/p_i$.  Thus,   the    expression  for
$\bar{t}_{0,N}$ is obtained in the closed form as
\eqb\label{eq:mfpt_1_app}
\bar{t}_{0,N} = N + \sum _{k = 0}^{N - 1} \omega_k + \sum _{k = 0}^{N - 2} \sum _{i = k + 1}^{N - 1} \left( 1 + \omega_k \right) \prod_{j = k + 1}^{i} \omega_j. 
\eqe

\section{Potential Profile Generation}\label{app2}

Given the pseudoenergy partition function 
\eqb
Z(\lambda) =  \int\mathcal{D}[U] e^{-\lambda\mathcal{H}[U]},
\eqe
the average pseudoenergy is
\eqb
\langle\mathcal{H}\rangle = \left.
-\frac{\partial}{\partial\lambda}\ln Z(\lambda)\right|_{\lambda = 1},
\eqe
and the variance is
\eqb
\langle(\Delta\mathcal{H})^2\rangle = 
\langle\mathcal{H}^2\rangle  - \langle\mathcal{H}\rangle^2 = \left.
\frac{\partial^2}{\partial\lambda^2}\ln Z(\lambda)\right|_{\lambda = 1}.
\eqe
Straightforward calculation for the pseudoenergy given by
Eq. (\ref{eq:pseudo}) yields
\eqb
\langle\mathcal{H}\rangle = L/2, \qquad 
\langle(\Delta\mathcal{H})^2\rangle = L/2.
\eqe

\begin{figure}[htb]
\includegraphics[width = 3.0in]{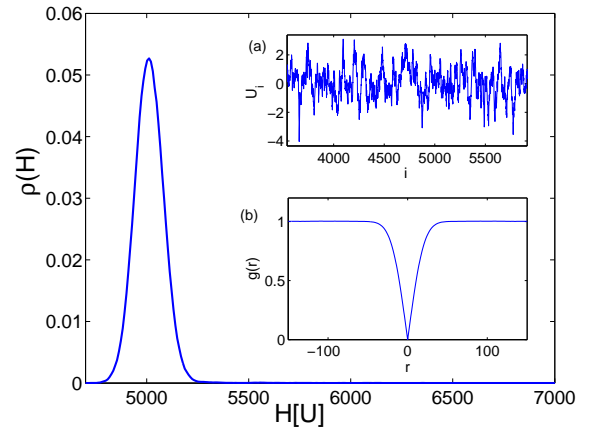}
\caption{\label{fig:sim_20}  Pseudoenergy  probability density for
a profile of length $L = 10000$, with $\sigma = 1.0$, $\xi_c =
20.0$. Insets: (a) Typical potential profile; (b) Potential profile
correllator $g(r) = 1/2\langle [U(x) - U(x+r)]^2 \rangle$; the
averaging was performed over 1000 profile realizations.}
\end{figure}

Hence, typical potential profiles have pseudoenergies in the range
$L/2\pm\sqrt{L/2}$. This result together with Gaussian statistics of
energy levels of Eq.~(\ref{eq:char_fun}) forms the basis of the algorithm we
employ for building the energy profiles.
First, a random and uncorrelated potential profile
obeying Gaussian statistics with the required variance $\sigma^2$
is generated on a one-dimensional lattice. Next, we look for a
permutation of lattice sites that produces a typical pseudoenergy
$\mathcal{H}[U]$ for a given correlation length $\xi_c$ (or,
equivalently, for given values of $\alpha$ and $\gamma$). This is
accomplished by a Metropolis-type algorithm that converges to a
prescribed value of pseudoenergy picked at random from Gaussian
distribution around $\langle\mathcal{H}\rangle$; see
Fig.~\ref{fig:sim_20}.

\bibliography{diff_corr_fin}

\end{document}